\documentstyle[12pt]{article}

\begin{document}

\baselineskip=20pt

\begin{center}
{\large \bf TIME OPERATOR FOR A QUANTUM SINGULAR OSCILLATOR}\\
\vspace{1.5cm}
{\normalsize \bf M.Martinis and V. Mikuta}\\
\vspace{0.5cm}
Division of Theoretical Physics,\\
Rudjer Bo\v skovi\'c Institute \\
10002 Zagreb, Croatia \\
\vspace{2cm}
{\large \bf Abstract}
\end{center}
\vspace{0.5cm}
\baselineskip=20pt

The problem of existence of a self-adjoint time operator conjugate
to a Hamiltonian with SU(1,1) dynamical symmetry is investigated.
In the space spanned by the eigenstates of the generator $K_3$ of
the SU(1,1) group, the time operator for the quantum singular
harmonic potential of the form $\omega ^2x^2 + g/x^2$ is
constructed explicitly, and shown  that it is related to the
time-of-arrival operator of Aharonov and Bohm. Our construction is
fully algebraic, involving only the generators of the SU(1,1)
group.

% The one-dimensional motion of aparticle in a singular harmonic
%potential of the form $\omega^2x^2 + g/x^2$ is studied.
%Using various nonlinear relations between the generators of the
%SU(1,1) group,

\vspace{1cm}
\newpage
\baselineskip=24pt

The unequal role played by time as an observable in classical and
quantum mechanics has been the source of controversy since the
early days of quantum mechanics. The problem arises because we
expect observables to be represented in quantum mechanics by
self-adjoint operators. However, a well-known argument due to
Pauli [1] stated that a self-adjoint time operator $T$ conjugate
to a self-adjoint Hamiltonian $H$ could not  be constructed if the
spectrum of $H$ is bounded from below.
% [1] W. Pauli: Handbuch
%derPhysik,ed.S.Fluege,vol.V/1,Springer-Verlag (1958)]

 Since then, the search for various time operators and the analysis of their
 self-adjointness and associated time-energy uncertainty relations
 have been the subject of a number of works [2]. The general concensus
 is that no such operator exists.
 %[2] J. Muga et al. Ann.Phys.\textbf{240},351 (1995);J.Muga et al.
 %Phys. Rev. A\textbf{58},4336 (1998)
 %L. Lynch, Phys. Rep. \textbf{256},367 (1995)
 Recently, the validity of  Pauli's objections has been
 critically evaluated [3], with the conclusion that there is no a
 priori reason to exclude the existence of  self-adjoint time
 operators canonically conjugate to a semibounded Hamiltonian. For
 this and other similar reasons, it seems reasonable to
 investigate explicit constructions of  time operators
 for various quantum mechanical systems.
 %[3] E.A. Galapon /quant-ph/9908033 v3

In this letter we pose a general problem of finding an operator
conjugate to a Hamiltonian with SU(1,1) dynamical symmetry. We
assume that the Hamiltonian is  linear in the generators $K_1 ,
K_2 ,K_3 $ of the su(1,1) algebra
\begin{equation}
\tilde{H} = \Omega _3 K_3 + \Omega_2 K_2 + \Omega_1 K_1,
\end{equation}
where the (2+1)- dimensional constant vector $\vec{\Omega } \equiv
(\Omega _1, \Omega_2, \Omega_3)$ has the norm $\Omega ^2 =
\Omega_{3}^2 - \Omega_{2}^2 - \Omega_{1}^2$. The  group generators
$K_3$ and $K_{\pm} = K_1 \pm iK_2$ satisfy the commutation
relations of the su(1,1) algebra:
\begin{equation}
[K_3 , K_{\pm}] = \pm K_{\pm},  \,\,\,\, [K_-, K_+ ] = 2K_3.
\end{equation}
Our objective here is to construct an operator $\tilde{T}$ in
terms of the generators $K_3,K_{\pm}$ that is  conjugate to the
Hamiltonian $\tilde{H}$ and satisfies $[\tilde{H},\tilde{T}] = i$.

In the following we  use the standard complete orthonormal basis
states $|n,k>$ that diagonalize the compact generator $K_3$. These
states are obtained from $|0,k>$ by n-fold application of $K_{+}$:
\begin{eqnarray}
|n,k> & = & \sqrt{\frac{\Gamma (2k)}{\Gamma (2k+n)n!}} (K_{+})^n
|0,k>,
 \nonumber\\ K_{-}|0,k> & = & 0,\\
K_3 |n,k> & = & (n+k)|n,k>,\,\, n = 0,1,2,... \nonumber
\end{eqnarray}
 The Bargman index k is related to the eigenvalue k(k-1) of the
 quadratic  Casimir operator $\hat{C} = K_{3}^2 - K_{1}^2 - K_{2}^2$.

 We  also need the Barut-Girardello  coherent states [4], which are
 the  eigenstates of $K_-$:
\begin{eqnarray}
K_-|z,k> & = & z|z,k>, \nonumber\\
|z,k> & = & \sum_{n=0}^{\infty}z^n \sqrt{\frac{\Gamma (2k)}{\Gamma
(2k+n)n!}}|n,k>,
\end{eqnarray}
where $z$ is an arbitrary complex number. These coherent states
can also be written as an exponential operator acting on the
vacuum state of $K_-$
\begin{equation}
|z,k> = e^{zK_+ (K_3 +k)^{-1}}|0,k>.
\end{equation}
In deriving this expression, we have used an operator identity\\
\begin{equation}
[K_+ (K_3 +k)^{-1}]^n = K_+^{n} \frac{\Gamma (K_3 +k)}{\Gamma (K_3
+k +n)}.
\end{equation}
Note also that the operator $K_+ (K_3 +k)^{-1}$ is canonical to
$K_-$:
\begin{equation}
[K_-,K_+ (K_3 + k)^{-1}] = 1.
\end{equation}

 The eigenvalue problem [5]  for our model Hamiltonian $\tilde{H}$ ,
\begin{eqnarray}
\tilde{H}|\Psi (\lambda )> & = & \lambda |\Psi (\lambda
)>,\nonumber\\
|\Psi (\lambda )> & = & \sum_{n=0}^{\infty}C_{n}(\lambda )|n,k>,
\end{eqnarray}
 depends on the choice of the vector $\vec{\Omega }$.\\ We   consider
 two cases [6]:\\
a) $\Omega ^2 >0, \,\,\, \Omega_3 >0$ when $\tilde{H}$ can be
transformed by means of the unitary operator to a standard form $H
= U^{\dagger}\tilde{H}U = \Omega K_3$.
The energy spectrum is discrete and bounded from below;\\
b) $\Omega ^2 = 0, \,\,\, \Omega_3>0$ when $\tilde{H}$ can be
transformed to \\$H = U^{\dagger}\tilde{H}U = \Omega_3 (K_3 -
K_1)$. In this case, the energy spectrum  is continuous and
bounded from below.

Let us  first consider the  time-operator problem  for a particle
moving in a repulsive singular potential of the Calogero-Moser
type [7]. The motion is described by the Hamiltonian
%F. Calogero, J. Math. Phys. {\bf 12}, 419 (1971); J. Moser, Adv. Math. {\bf 16},1 (1975)
\begin{equation}
H = \frac{1}{2} ( p^2 + \frac{g}{x^2}), \,\, g > 0.
\end{equation}
This Hamiltonian is interesting for several reasons:\\ i) It is
scale invariant and  has the full conformal group  as a dynamical
symmetry group [8] with the generators $H$, $D = -(xp + px)/4$,
and $K = \frac{1}{2}x^2$, which obey the algebra
\begin{equation}
[H,D] = iH, \;   [K,D] = -iK, \;   [H,K] =  2iD
\end{equation}

with a constant Casimir operator $\hat{C} = \frac{1}{2}(HK + KH) -
D^2 = \frac{g}{4} - \frac{3}{16}$.\\ ii)The spectrum of $H$ is
positive, continuous, and bounded from below, with a
non-normalizable ground state [8 ].\\iii) It can  be easily
extended to the well-known one-dimensional N-body problem of Calogero-Moser [7].\\
iv) Recently, it has been observed that the dynamics of particles
near the horizon of a black hole is also
associated with this Hamiltonian [9].\\

If we now identify
\begin{eqnarray}
K_1 & \equiv S & =  \frac{1}{2}(\omega K -
\frac{1}{\omega}H),\nonumber\\
K_2 & = & D, \\
K_3 & \equiv R & = \frac{1}{2}(\omega K +
\frac{1}{\omega}H),\nonumber
\end{eqnarray}
it can be seen that the conformal algebra (10) is isomorphic to
the algebra of $SU(1,1)\sim O(2,1)\sim SL(2,\cal R)$ with the
Bargman index $k = \frac{1}{2}(1 + \sqrt{g + \frac{1}{4}})$.
 We note that $H = \omega (K_3 - K_1)$ and $\omega K = K_3 + K_1$
 are  related to $K_-$ as follows :
\begin{equation}
e^{-\omega K} H e^{\omega K} = -2\omega K_-.
\end{equation}
The energy eigenstates of $H|E> = E|E>$ are thus seen to be
proportional to the  Barut-Girardello coherent states [4,10] with
$z = -E/2\omega $:
\begin{equation}
|E> = e^{\omega K}| - \frac{E}{2\omega},k>.
\end{equation}
Note that the eigenstate $<x|E>$ in the limit $E\rightarrow 0$ is
not normalizable, since lim$_{E\rightarrow 0}<x|E> = <x|e^{\omega
K}|0,k> \propto \omega^k x^{2k-1/2}$. \\The dificulty arises from
the oscillating behavior of $<x|E>$ at large distances [8].

Combining the relations (7) and (12), we find that the operator
\begin{equation}
T(\omega ) = -\frac{i}{2\omega }e^{\omega K} K_+ (K_3 +k)^{-1}
e^{-\omega K},\;\; T^{\dagger}(\omega) = T(-\omega)
\end{equation}
has the property  $[H,T(\omega )] = i$ and can be interpreted as a
possible time operator conjugate to $H$. Since $\omega $ is a free
parameter, $T(\omega )$ generates an uncountable number of
different time  operators canonically conjugate to $H$. In the
limit $\omega \rightarrow 0$, we find
\begin{equation}
T(\omega ) \rightarrow  \frac{i}{2\omega } + \frac{1}{\sqrt{H}} D
\frac{1}{\sqrt{H}}+ i\frac{2k+1}{2H} + \cal{O}(\omega ).
\end{equation}
It is important to point out here that the solution of $[H,T]=i$
is not unique. Any $\bar{T} = T + \phi (H)$, with arbitrary
$\phi$, satisfies the same canonical commutation relation.
Therefore, using the concept of a minimal solution we  choose a
Hermitian version of $T(\omega)$ as a time operator conjugate to
$H$, which in the limit  $\omega \rightarrow 0$ becomes
\begin{eqnarray}
T & = & lim_{\omega \rightarrow 0}(T(\omega ) + T^{\dagger}(\omega
))/2 \nonumber\\
& = & \frac{1}{\sqrt{H}} D \frac{1}{\sqrt{H}}.
\end{eqnarray}

 We  argue  that $[H,D] = iH$ is the key relation [11,12] for defining the time
 operator. In fact, requiring that
 $\sqrt{H}T\sqrt{H} = D$, we can immediately deduce that the commutators of $H$ and $T$
 must have the following form:
 \begin{equation}
 [H,T] = i(1+X),
\end{equation}
where the operator $X$ is such that $HXH = 0$.
 In the limit $g\rightarrow 0$, we have $H\rightarrow H_0$ and
$T\rightarrow T_0$, so that
\begin{equation}
 [H_0,T_0] = i(1+X_0), \,\,\,\, H_0X_0H_0 = 0,
\end{equation}
where $H_0 = p^2/2$ and
\begin{eqnarray}
T_0 & = & \frac{1}{\sqrt{H_0}} D \frac{1}{\sqrt{H_0}}\nonumber\\
& = & - \frac{1}{2}(x\frac{1}{p} + \frac{1}{p}x)
\end{eqnarray}
is the time-of-arrival operator of Aharonov and Bohm [13].The
operators $T$ and $T_0$ can also be related to each other   by
means of a unitary operator [14] that transforms $H \rightarrow
H_0$,
 so that we obtain
\begin{eqnarray}
H  & = &  UH_0 U^{\dagger }, \nonumber\\
T  & = &  UT_0U^{\dagger },\\
U & = & e^{-i\pi K_3} e^{i\pi K_{3}^{0}}, \nonumber
\end{eqnarray}
where $ K_{3}^0 = K_3(g=0)$.

 Finally, we consider the quantum singular harmonic
 oscillator of the Calogero-Sudarshan type [15], which
 is  proportional to  the $K_3$
 generator of the $SU(1,1)$ group:
 \begin{equation}
 H_{CS} = 2\omega K_3.
 \end{equation}
 To construct the time operator for $H_{CS}$, we first observe the
 relationship [16] between $H_{CS}$ and the Hamiltonian for the ordinary
 harmonic oscillator, $H_h = H_0 + \omega^2 K = H_{CS}(g=0)$ :
\begin{eqnarray}
H_{CS} & = &  U_1 H_hU_{1}^{-1}, \nonumber\\
U_1 & = & e^{-K_-} e^{K_{-}^{0}},
\end{eqnarray}
where $K_{-}^{0} = K_{-}(g=0)$. The time operator for $H_h$ was
 constructed and discussed earlier in [12,17]. Its construction is
 simple if we observe that the Casimir operator with $k = 3/4$ can be
 used to express the operator $K$ in the form
\begin{eqnarray}
K & = & T_0 H_0 T_0 + \frac{1}{16H_0} \nonumber \\
& = & QH_0Q - \frac{i}{2}Q, \\
Q & = & - T_0 + \frac{i}{4H_0}.\nonumber
\end{eqnarray}
Then the Hermitian operator
\begin{equation}
T_h = \frac{1}{2}(T_h(Q) + T_h^{\dagger }(Q))
\end{equation}
satisfies $[H_h,T_h] = i$, where
\begin{equation}
T_h(Q) = \frac{1}{\omega }arctg (\omega Q).
\end{equation}
It is now easy to see that the time operator for the Hamiltonian
$H_{CS}$ is
\begin{equation}
T_{CS} = U_1 T_h U_1^{-1}.
\end{equation}

In conclusion, we have presented an algebraic method of
constructing  Hermitian operators conjugate to a Hamiltonian with
$SU(1,1)$ dynamical symmetry. The time operator for the quantum
singular harmonic potential is constructed explicitly and shown
that it is related to the time-of-arrival operator, $T_0$ of
Aharonov and Bohm. The question whether  time operators thus
constructed are self-adjoint operators in Hilbert space requires a
careful examination of their spectra and eigenfunctions. The
eigenvalue problem of the operator $T_0$ can be solved in momentum
space [2,18]. It is not self-adjoint and its eigenfunctions are
not orthogonal. The same conclusion can be reached for the time
operator $T$ owing to the relation (20). For $T_h$ and $T_{CS}$
this problem is still open [19,20].

This work was supported by the Ministry of Science and Technology
of the Republic of Croatia under Contract No.0098004.

\end{document}